\documentclass[epj]{svjour}

\usepackage{graphicx}	
\usepackage{dcolumn}	
\usepackage{bm}			

\begin{document}

\title{ {\it Ab initio\/} calculations of magnetic structure
         and lattice dynamics of Fe/Pt multilayers }

\author{ Pawe\l{} T. Jochym\inst{1,}\thanks{{\tt e-mail: Pawel.Jochym@ifj.edu.pl}} \and 
                Krzysztof Parlinski\inst{1} \and 
                Andrzej M. Ole\'{s}\inst{1, 2}
}
\institute{ Institute of Nuclear Physics, Polish Academy of Sciences, 
              Radzikowskiego 152, PL-31342 Krak\'{o}w, Poland 
              \and
              Max-Planck-Institut f\"ur Festk\"orperforschung, 
	      Heisenbergstrasse 1, D-70569 Stuttgart, Germany 
}

\date{\today}

\abstract{
The magnetization distribution, its energetic characterization by the 
interlayer coupling constants and lattice dynamics of (001)-oriented
Fe/Pt multilayers are investigated using density functional theory 
combined with the direct method to determine phonon frequencies. 
It is found that ferromagnetic order between consecutive Fe layers is 
favoured, with the enhanced magnetic moments at the 
interface. The bilinear and biquadratic coupling coefficients between Fe 
layers are shown to saturate fast with increasing thickness of 
nonmagnetic Pt layers which separate them. The phonon calculations 
demonstrate a rather strong dependence of partial iron phonon 
densities of states  on the actual position of Fe monolayer 
in the multilayer structure.
}


\maketitle

\section{Introduction}
\label{intro}

The technological importance of metallic multilayers consisting of 
alternating layers of magnetic and nonmagnetic material stimulated 
both experimental and theoretical studies of these systems which 
continues over the last two decades. The properties of magnetic 
multilayers are to a large extent determined by the interface properties. 
In order to understand whether future applications of multilayers as 
magnetic data storage materials are possible, their microscopic 
properties have to be investigated to determine the magnetic 
interactions on the nanoscale and to understand the dependence of these
interactions on the multilayer structure. Since the discovery of 
composition and annealing dependence of the magnetic properties in Fe/Pt 
bilayers \cite{Kit87}, a number of experimental papers have been devoted 
to the magnetic properties of the Fe/Pt multilayers \cite{Sakur,Sim96}.

The materials themselves are of great technological importance since 
they are characterized by large uniaxial magnetocrystaline anisotropy 
energy \cite{Sak94}, which in turn leads to high temperature stability 
of the data magnetically recorded in such a medium \cite{Mos02}.
The magnetocrystaline anisotropy energy density has been already 
analysed \cite{Daa91,Sak94} using {\it ab initio\/} techniques. However, 
to the best of our knowledge, the dependence of the type of magnetic 
ordering on the thickness of Fe and Pt layers, as well as lattice 
dynamics in these systems, have not been investigated with modern 
computational techniques until now. 

The purpose of the present paper is to analyse magnetic ordering and 
lattice dynamics in the (001)-oriented Fe/Pt multilayers using an 
{\it ab initio\/} approach. 
We have computed the magnetization distribution in a series 
of representative multilayer structures with one up to twelve atomic
monolayers (MLs) in an iron layer, and one up to five platinum MLs in 
the spacer layer. The calculations were performed self-consistently and 
the configurations with ferromagnetic (FM) and antiferromagnetic (AF) 
order between the consecutive iron layers were obtained and their 
energies and magnetization distributions were compared 
with each other. We also report the values of bilinear and biquadratic 
coupling constants, which quantify the energy difference between the 
AF and FM configuration of the Fe layers within the model introduced 
by Fullerton and Bader \cite{Ful96}, and depend on the thickness of  
both magnetic iron layers and (almost) nonmagnetic platinum layers.

The paper is organized as follows. The computational methods are shortly
described in Sec. II. In Sec. III we present the magnetic properties
of Fe/Pt multilayers, and systematically investigate the energy difference 
between the AF and FM configurations, depending on the number of Fe and Pt 
atomic MLs in the considered multilayer structure. The phonon dispersions 
and the phonon densities of states for two representative Fe/Pt multilayer 
systems are analysed in Sec. IV. We present a short summary and the main
conclusions in Sec. V.

\section{Computational methods}
\label{methods}

The calculations were carried out with standard density functional 
theory (DFT) code as implemented in the {\sc vasp} package \cite{Kre96}.
We have used well established ultrasoft generalized gradient 
approximation (GGA) pseudopotentials \cite{Kre99} with Perdew--Wang 91
gradient correction scheme (as given in Ref. \cite{Per92}) as input for 
phonon calculations. The calculations of magnetic structure of the 
system were carried out with pseudopotentials introduced by Perdew, 
Burke and Ernzerhof \cite{Per96,Blo94} (so-called PBE-GGA pseudopotentials) 
as implemented in the {\sc vasp} package \cite{Kre99}. The difference 
in method used for these two calculations does not influence our 
general conclusions since the results from different methods are not 
compared with each other. The calculations neglected spin-orbit 
coupling which may somewhat influence the results, especially for Pt 
atoms, since they possess fairly strong spin-orbit component.

\begin{figure}[t!] 
\centering 
\includegraphics[width=7.7cm]{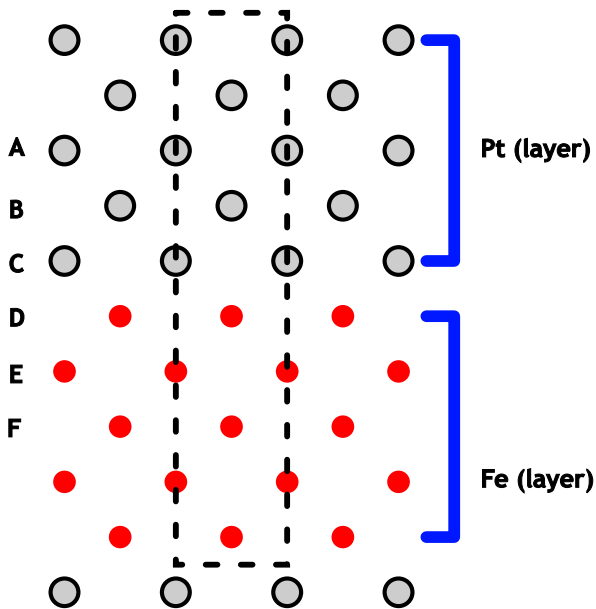} 
\caption{
Side view of a representative 5/5 Fe/Pt multilayer structure before
geometry optimization
(with five atomic MLs of iron and five atomic MLs of platinum). 
Non-equivalent MLs in the unit cell are labeled by characters $A-F$; 
the unit cell in the FM structure, used also for phonon calculations, 
is indicated by dashed lines 
(the magnetic unit cell in an AF configuration is twice larger). 
The bottom row of Pt atoms is shown only for clarity.
}
\label{fig:structure}
\end{figure}

The series of structures investigated in this paper was derived from 
the basic (001) bcc iron lattice with platinum atoms replacing iron 
atoms within platinum layer (by layer we mean one or more MLs of the same 
material). In this way some MLs of Fe atoms were replaced by Pt ones. 
Then the entire multilayer structure was optimized with respect to 
atomic positions and its unit cell size. The convergence criterion 
has been set at root mean square (RMS) force $\leq 10^{-7}$ eV/\AA{} 
and RMS stress $\leq 10$ MPa. The atomic forces and magnetic moment 
density distribution have been calculated using final optimized 
structures. One example of such a structure before optimization is
shown in Fig.~\ref{fig:structure}; its unit cell in the FM (ground state) 
configuration contains atoms belonging to five atomic MLs of iron and 
to five atomic MLs of platinum, in so-called 5/5 system.

\begin{figure}[t!] 
\centering 
\includegraphics[width=8cm]{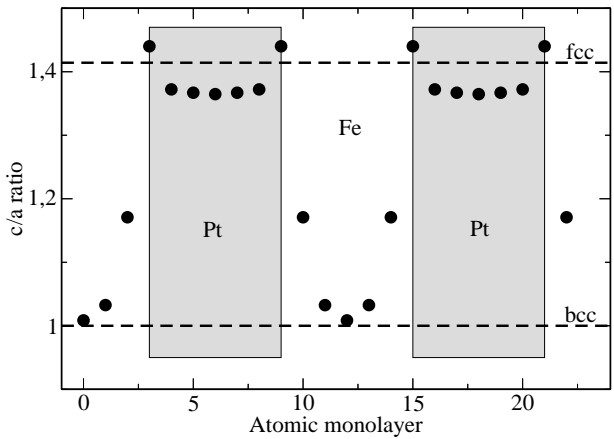} 
\caption{
Inter-layer distance in the 4/8 structure after structure
optimization. Dashed lines indicate c/a ratio for ideal bcc (bottom
line) and fcc (top line) structures. The shaded areas indicate the
extension of the platinum layers.}
\label{fig:layerdist}
\end{figure}

The structure optimization led to substantial changes in the structure 
of the considered multilayer systems. Namely, the inter-layer distance
in the Pt layer increased to the point where this layer adopts the well
known experimentally fcc structure. This change is illustrated in 
Fig.~\ref{fig:layerdist}, showing the local $c/a$ ratio of lattice constants 
for each atomic monolayer in the structure. It is straightforward to see by 
analysing these data that the structure of the platinum layer changes from 
the starting bcc structure (assumed at the beginning of the optimization 
process) to the fcc lattice -- well-known experimentally. This may be 
interpreted, together with phonon calculations below, as a strong 
evidence of rather realistic description of the physical properties of 
the multilayer systems in the presented calculations. The above change 
starts to show up for quite thin Pt layers --- even the system consisting
of four Pt atomic MLs shows already the interlayer spacing close to the 
fcc structure. For thicker Pt layers this effect is very strong (see
Fig.~\ref{fig:layerdist}). The above finding agrees quite well with
experimental findings of Antel {\em et al.} \cite{Antel99}, where a
similar process of gradual change from tetragonally distorted bcc 
(bct) to tetragonally distorted fcc (fct) lattice was observed. The 
iron layer is quite stable but when it is thin (few atomic MLs) it is 
prone to be distorted towards the fcc structure. This tendency may be 
also observed in Fig.~\ref{fig:layerdist}, where the distortion of 
$c/a$ ratio goes deep into the Fe layer, while almost the whole Pt 
layer keeps its optimum $c/a$ ratio. Similar structural changes in 
Fe/Pt systems have also been observed experimentally \cite{Antel99,Sakur}.

All calculations were carried out for spin-polarized systems with the 
self-consistent optimization of the magnetization distribution within 
a given Fe/Pt multilayer system. The set of calculated structures 
contains a range of variants from the simplest 1/1 structure (periodic 
structure with 1 ML of Fe and 1 ML of Pt in its unit cell) up to 4/8 
(i.e., with 4 atomic MLs of Fe and 8 atomic MLs of Pt) in increments of 
2 MLs, and also, in addition, 1/12 structure 
(1 ML of Fe and 12 MLs of Pt). 

The magnetic properties of Fe/Pt multilayers were investigated 
systematically (see Sec. III) by changing the thickness of Fe and Pt 
layers in a periodic system consisting of four layers, i.e., twice 
larger than that shown in Fig.~\ref{fig:structure}, with two Fe and two 
Pt spacer layers interchanged with each other, with periodic 
boundary conditions. Several different starting configurations for the 
magnetization distribution were used in all cases in order to obtain 
unbiased results for a given system. Here we have used a similar approach 
to that employed to investigate Fe/FeSi multilayers in our recent paper
\cite{Joc06}.

Starting from equilibrium multilayer configurations obtained as a   
result of electronic structure calculations, we investigated 
lattice dynamics for two representative Fe/Pt multilayers: 4/4 and 5/5
systems (see Fig.~\ref{fig:structure}), reported in Sec. IV. Thereby we
used the so-called direct method of lattice dynamics calculation 
\cite{Par97}. It follows the early idea of calculating the force 
constants with {\it ab initio\/} methods \cite{Kun82} by implementing the
full crystal symmetry given by the crystallographic space group 
\cite{Par97}. As such, it is able to achieve high accuracy and became 
a very powerful computational tool in the last decade. By construction
it allows one to determine accurate phonon frequencies at the high 
symmetry points in the reciprocal space, and gives otherwise the phonon 
spectra with high accuracy, if the force constants decrease fast enough 
with increasing distance, as in the present case. 

The formulation of Ref. \cite{Par97} provides an algorithm to 
calculate force constants automatically for a system of any size. 
The procedure employed in the present study used displacements of 
0.03~\AA{} in each non-equivalent direction for every non-equivalent atom. 
The number of necessary displacements was thus determined by the system 
symmetry and by the size of the multilayer unit cell. Next, the matrices 
of force constants were found using the singular value decomposition 
method. Finally, the dynamical matrix was constructed and diagonalized 
for each wave vector~${\bf k}$.

\section{Magnetic properties}
\label{magnetic}

\begin{figure}[t!] \centering 
  \includegraphics[width=8cm]{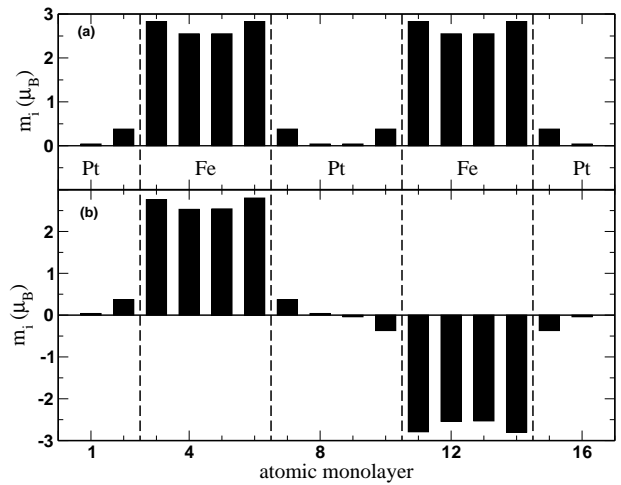}
\caption{
Distribution of magnetic moments $m_i$ in the 4/4 multilayer structure 
consisting of two layers with four Fe MLs separated by Pt spacer with 
four Pt MLs, as obtained for two magnetic configurations:
(a) two magnetic unit cells in the FM ground state, and 
(b) magnetic unit cell of the (locally stable) AF order between two 
neighbouring Fe layers.}
\label{fig:magprofile}
\end{figure}

When Fe layers are separated by layers of nonmagnetic Pt spacer in a
Fe/Pt multilayer, the first question concerns the distribution of 
magnetization over the Fe atoms. In the case of platinum spacer we 
have found that the interface proximity substantially enhances 
magnetic moments of iron atoms. This can be clearly observed for the 
4/4 Fe/Pt structure, with magnetic moments $m\sim 2.8$ $\mu_B$ and 
$\sim 2.5$ $\mu_B$ for the Fe atoms at the interface and in the inner 
Fe ML (Fig.~\ref{fig:magprofile}), respectively, and with differences 
less than 2\% between the FM and AF configuration. 

Also in larger systems we have found that the magnetic moments $m$ at 
the interface are systematically {\it enhanced\/} and are rather close 
to 3.0 $\mu_B$, i.e., the $3d$ electrons are almost fully polarized. 
In contrast, the moments at the Fe atoms of inner MLs are smaller and
gradually approach the bulk value, $m_{\rm Fe}\simeq 2.2$ $\mu_B$, with
increasing thickness of the Fe layer. While the later behavior may be 
expected and in fact is observed experimentally \cite{Antel99}, the 
magnetization increase at the interface Fe atoms characterized by a lower 
coordination of Fe atoms is not obvious. We interpret it as a manifestation 
of electron correlation effects which increase local moments at Fe atoms 
when the kinetic energy of $3d$ electrons is decreased by a lower Fe 
coordination \cite{Ole84}. In agreement with this interpretation, the 
moment enhancement at the interface becomes even more pronounced in the 
extreme case of a single iron ML, where Fe atoms have only platinum 
coordination and their magnetic moments are increased up to the level 
of saturation $m\simeq 3.0$ $\mu_B$, i.e., by 35\% over the bulk iron 
value $m_{\rm Fe}\simeq 2.2$ $\mu_B$ (the magnetic moments of Pt atoms 
at the interface are here quite small $m_{\rm Pt}\simeq 0.3$ $\mu_B$).

The above magnetic moment amplification effect at the interface in 
case of Pt spacer is remarkably different from the behaviour found 
recently for Fe/FeSi multilayers \cite{Joc06}. There, the magnetic 
polarization at interface Fe atoms is {\it reduced\/}, while the 
magnetic moments at Fe atoms of the second ML are somewhat 
{\it enhanced\/}. This demonstrates, that the moment enhancement may
occur only in case of a metallic spacer, while we suggest that the 
electronic states at an insulating interface behave more like molecular
bonding and thus $3d$ electrons are partly bound and the magnetization.
is reduced.

In addition, large magnetic moments at Fe atoms of the metallic 
interface induce weak FM polarization in the neighbouring Pt ML 
($m\simeq 0.4$ $\mu_B$ in case of 4/4 structure of 
Fig.~\ref{fig:magprofile}). The induced magnetization of the Pt layer has 
been observed experimentally \cite{Antel99} to be on the same level as 
calculated here. There is however a difference in the distribution of 
the magnetic moment inside Pt layer --- the present results indicate 
large enhancement at the interface, while the experimental results 
suggest more uniform distribution of the moment across the whole Pt 
layer. This difference may be connected with the contribution of 
spin-orbit interaction which, according to the experimental results 
\cite{Antel99}, is absent or less pronounced in the case of 
iron-transition metal multilayers. It might be also to certain extent
attributed to interdiffusion in the measured sample, following the 
suggestion of Ref.~\cite{Antel99}. Unfortunately, such effects cannot 
be included in the electronic structure calculations at present.

Our main goal in investigating the magnetic properties of multilayers 
is to establish the dependence of the favoured magnetic ordering in the 
underlying system on the thickness of the spacer (here Pt) layer and 
to characterize it by certain global parameters. Here we follow 
a phenomenological model of Fullerton and Bader \cite{Ful96} introduced 
to describe the energy difference between an AF and a FM configuration
of a given multilayer structure. In this model it is assumed that the 
energy difference between an AF and a FM configuration of the system, 
\begin{equation}
\Delta E=E_{\rm AF}-E_{\rm FM}, 
\label{eex}
\end{equation}
may be expressed by the following simple formula:
\begin{equation}
E_{\rm AF(FM)}= J_1\, {\bf m}_1\cdot{\bf m}_2 
        + J_2\,({\bf m}_1\cdot{\bf m}_2)^2 + E_0 + E_m.
\label{ej}
\end{equation}
Here $J_1$ and $J_2$ are the bilinear and biquadratic coupling 
parameters, respectively, while ${\bf m}_1$ and ${\bf m}_2$ are 
magnetic moments of two interacting layers normalized per area unit 
(only the interactions between the nearest neighbour magnetic layers are 
considered). $E_0$ and $E_m$ are reference energies of nonmagnetic
structure and magnetic energy of the isolated layers, respectively. 
Note that the magnetic moments depend on the number of Fe MLs, and are 
practically the same for the AF and the FM configuration of a given 
multilayer structure (see below).

The derivation of the parameters $\{J_1,J_2\}$ in Eq. (\ref{ej}) 
requires the calculation of energies of the (locally stable) AF, FM and 
nonmagnetic variants of each investigated structure separately. Here 
we use the same method as the one introduced recently for Fe/FeSi
multilayers \cite{Joc06}. Note that the same result could also be 
obtained by calculations performed for two systems with spins oriented 
in two different directions (e.g. the reference system and the one 
rotated by 90 degrees), but this would require employing the non-collinear
spins technique which is more involved and more technically demanding 
than the approach used here. In each case we determined these locally 
stable configurations and next analysed the results in order to 
extract information about the interlayer exchange couplings. 

Furthermore, the procedure used here requires the values of magnetic 
moments ${\bf m}_l$ of two coupled layers $l=1,2$. The magnetic moments 
have been derived from the calculated spin-polarized electronic structure 
by numerical integration of the magnetic moment density in the spherical 
volume with radius $r=1.3$ \AA{} around each atom. These individual 
moments were subsequently summed up within respective layers. We have 
verified that the resulting moments for each layer were practically 
the same (the difference is less than $\leq$ 1\%) for both AF and FM 
configuration of the same structure. Further details concerning this 
procedure of calculating the coupling parameters $\{J_1,J_2\}$ were 
described in the previous paper \cite{Joc06}.

\begin{figure}[t!] \centering 
  \includegraphics[width=7.5cm]{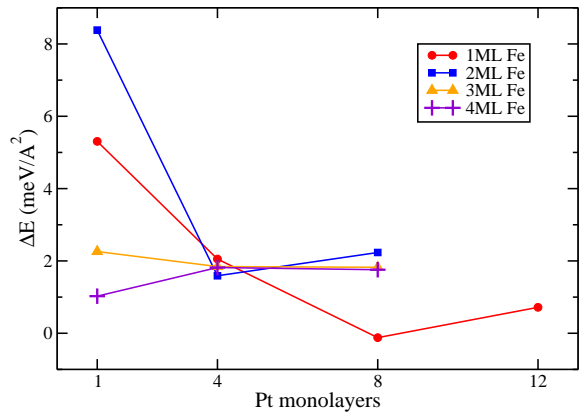}
\caption{
Energy difference $\Delta E$ (\ref{eex}) between AF and FM order of Fe 
layers in Fe/Pt multilayers for $1-4$ atomic Fe MLs, as a function of 
increasing thickness of the Pt spacer layer.}
\label{fig:deltaE}
\end{figure}

The energy difference $\Delta E$ (\ref{eex}) is presented in 
Fig.~\ref{fig:deltaE} as a function of the increasing thickness of platinum
spacer layer for several Fe/Pt multilayers. The data shows a clear 
preference towards the FM ordering of consecutive Fe layers already for 
the systems with only up to four Fe MLs. Note, however, that the energy 
difference $\Delta E$ decreases (with the exception of a single Fe ML)
with increasing thickness of the iron layer, while the platinum layer 
consists of a single ML. 

The above behaviour is again different from the Fe/FeSi system, where 
the order between the Fe layers is AF for thin layers, and becomes FM 
only for sufficiently thick FeSi spacer. Furthermore, it is quite 
remarkable that the energy difference $\Delta E$ for all the considered 
Fe/Pt multilayers except for the system with one Fe ML converges quite 
fast to a constant value with the increasing thickness of the Pt layer.  
This feature appears to be again universal with exception of the system 
with a single iron ML which does not contain bulk-like iron atoms with 
only iron coordination, and thus is expected to behave in a different 
way. 

\begin{figure}[t!] \centering 
  \includegraphics[width=8cm]{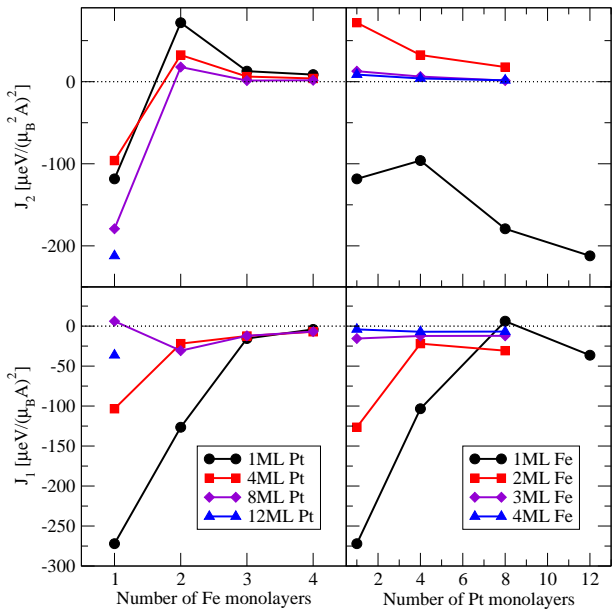}
\caption{
Bilinear $J_1$ (bottom) and biquadratic $J_2$ (top) coupling coefficients 
defined by Eq. (\ref{ej}) as obtained for Fe/Pt multilayers for 
increasing thickness of iron layer (left) and platinum spacer layer 
(right).}
\label{fig:j1j2}
\end{figure}

The thickness dependence of the coupling coefficients between Fe layers 
$J_1$ and $J_2$, see Eq. (\ref{ej}), are presented in Fig.~\ref{fig:j1j2}.
It is interesting to compare them with the ones
obtained before for metal-insulator Fe/FeSi multilayers. There are 
some similarities and some differences between the characters of their 
thickness dependence in Fe/FeSi multilayers \cite{Joc06}, and in the 
present metallic Fe/Pt ones. First, in both 
cases, one finds rather weak dependence of $J_2$ on the thickness of 
the spacer layer (top right panel of Fig.~\ref{fig:j1j2}). In case of 
one or two Fe MLs there is also no noticeable convergence of $J_2$ with 
increasing thickness of the spacer layer, while for thicker Fe layers
the values of $J_2$ are very small indeed for 8 Pt MLs and suggest that 
this parameter becomes gradually irrelevant for increasing thickness of 
spacer layers. Second, there is a qualitative difference in $J_2$ behaviour 
as a function of iron layer thickness with the one found in Fe/FeSi 
systems. In the case of Fe/FeSi multilayers one finds consistent and 
close to the exponential decay of $J_2$, while for Fe/Pt we could spot 
some kind of oscillatory shape in the $J_2$ plot (top left panel of 
Fig.~\ref{fig:j1j2}). This we interpret as a manifestation of the
metallic behaviour, in contrast to the gradual decay of the coupling
parameter $J_2$ in an insulating medium.

On the contrary, the behaviour of the bilinear coupling parameter $J_1$ 
(bottom part of Fig.~\ref{fig:j1j2}) shows more similarities between 
Fe/FeSi and Fe/Pt case. In both systems the values of the $J_1$ 
converge quite fast with increasing thickness of both iron and spacer 
layer. In case of the iron-thickness dependence (bottom left panel of
Fig.~\ref{fig:j1j2}) the asymptotic value appears to be close to zero, 
while the data suggest that it is finite for the spacer-thickness 
dependence (bottom right panel of Fig.~\ref{fig:j1j2}).

A single ML of iron is again remarkably different from other multilayer
systems with thicker Fe layers. 
The values of $J_2$ coupling constant are in this case {\it negative\/} 
and rather large, in contrast to thicker Fe layer. Furthermore, the 
values of $|J_2|$ do not decrease with increasing thickness of Pt
spacer, unlike for thicker layers. Also the values of $J_1$ show a 
different behaviour from that found for ticker layers, and the change 
to positive $J_1$ found at 8 Pt MLs followed by a large negative value 
for 12 Pt MLs is even somewhat surprising. It suggests that the 
magnetization density within Pt layers exhibits oscillations, which in 
some cases may even lead to weak AF coupling (see also 
Fig.~\ref{fig:deltaE}). Altogether, all these features confirm that the 
behaviour of a single Fe ML is dominated by the interface, and follows 
from a rather strong FM polarization of Pt atoms at the metallic 
interface, which extends into the Pt layers.

\section{Lattice dynamics}
\label{phonons}

We complete the discussion of the physical properties of Fe/Pt 
multilayers by presenting the phonon spectra for selected optimized 
structures. In order to reach unbiased conclusions, the lattice 
dynamics has been calculated for two systems, using the 
$2\sqrt{2}\times2\sqrt{2}\times 1$ supercell for the 4/4 multilayer
structure, and the $\sqrt{2}\times\sqrt{2}\times 1$ supercell for the 
5/5 one, respectively
(the side view of the 5/5 structure is shown in Fig.~\ref{fig:structure}). 
These calculations are rather demanding and do not allow us to treat 
more atoms in a unit cell at present, but the unit cells of these two 
systems are already large enough to identify certain generic features of 
the phonon spectra in the investigated family of Fe/Pt multilayers. To 
understand them better, we first summarize shortly the results of the 
electronic structure calculations --- they yield a tetragonal structure 
with $a=b=4.0246$ \AA{} and $c=13.7906$ \AA{} lattice constants. 
The former value is only slightly larger than iron lattice parameters 
(we found $a=4.0188$ \AA{} for a test calculation with the same setup). 
This means that there are only small stresses in the multilayer --- 
judging from the compressibility of the iron structure not higher than 
of the order of 0.3~GPa. The interlayer spacing results are the same
as described above in section~\ref{methods}.

\begin{figure}[t!] \centering 
  \includegraphics[width=8cm]{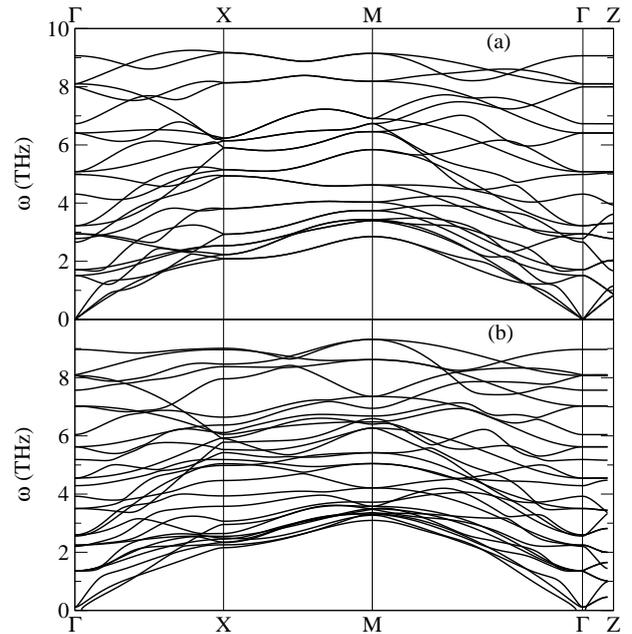}
\caption{
Phonon dispersion relations of the Fe/Pt multilayer structures in FM 
configuration as functions of wave vector ${\bf k}$ along high symmetry 
directions in the Brillouin zone for:
(a) 4/4 multilayer, and
(b) 5/5 multilayer of Fig.~\ref{fig:structure}. 
The high symmetry points in units of $\frac{2\pi}{a}$ are: 
$\Gamma=(0,0,0)$, $X=(1,0,0)$, $M=(1,1,0)$, $Z=(0,0,1)$.
}
\label{fig:dispcurv}
\end{figure}

The equilibrium configuration was determined by comparing the total 
energies of several possible stackings of Pt layer on Fe layer. The 
structure of the layers themselves was chosen following the experimental 
evidence that the bcc iron and fcc Pt are present in this material.
The crystal structure of the 4/4 system is given by the $P4/nmm$ space 
group, labelled as 129 (group $D_{4h}^7$ in Schoenflies notation), and 
for the 5/5 one it is $P4mm$, 99 ($C_{4v}^1$) respectively. 
We have verified that the structure with the lowest energy is the one 
with a straightforward continuation of lattice points from bcc lattice 
of iron to fcc lattice of platinum, as shown in Fig.~\ref{fig:structure} 
and described in section~\ref{methods}.

\begin{figure}[t!] \centering 
\includegraphics[width=8cm]{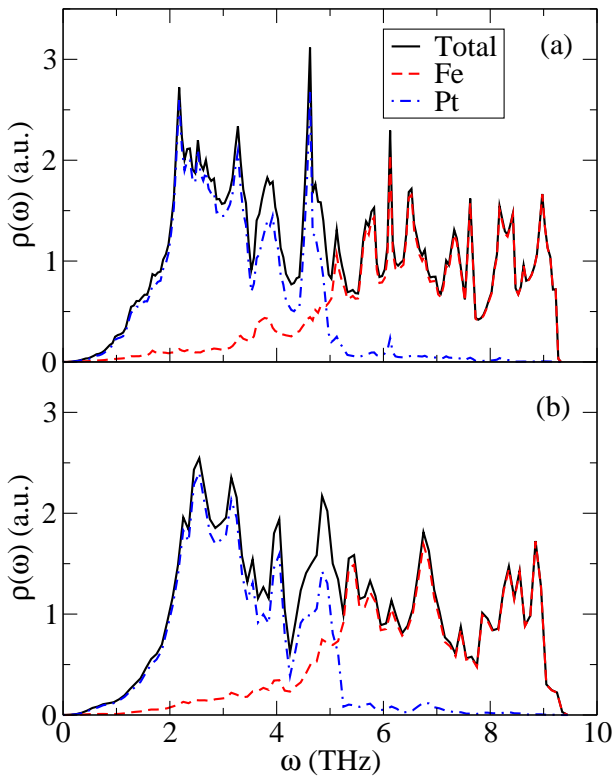}
\caption{
Total phonon density of states $\rho(\omega)$ (solid lines) and 
partial phonon densities of states for Fe (dashed lines, red) and Pt 
(dashed-dotted, blue) atoms, respectively, as obtained for the FM
ground state with:
(a) 4/4 multilayer, and
(b) 5/5 multilayer, represented schematically in Fig.~\ref{fig:structure}.}
\label{fig:phdos}
\end{figure}

The calculated phonon dispersion relations for the 4/4 and 5/5
structures are shown in Fig.~\ref{fig:dispcurv}. It is clear that the 
shape of the dispersion curves is quite similar, and the energy range
of phonon spectra is the same for both systems. The main difference is 
in the number of branches, due to the different symmetries of the 
investigated multilayers. Thus one finds 3 acoustic and 27 (21) optic 
phonon branches for the 5/5 (4/4) multilayer, and some differences in 
the detailed dispersions of individual phonon modes. 
However, in Fig.~\ref{fig:phdos} we can clearly see that these 
differences do not affect significantly the distribution of spectral 
weight in the total phonon density of states $\rho(\omega)$, and the
overall frequency ranges for the individual contributions to 
$\rho(\omega)$. This observation gives further support to the present
choice of the supercell --- even the $\sqrt{2}\times\sqrt{2}\times 1$ 
supercell used for the 5/5 system is large enough to obtain meaningful 
and representative results for the phonon density of states, while for 
detailed analysis of the dispersion curves it is necessary to consider 
instead a larger $2\sqrt{2}\times 2\sqrt{2}\times 1$ supercell, as in 
the 4/4 system.

The phonon density of states function $\rho(\omega)$ in 
Fig.~\ref{fig:phdos} consists of two distinct contributions --- heavier 
platinum atoms contribute with lower frequencies, while the vibrations 
of iron atoms are seen in the upper part of $\rho(\omega)$, well separated 
from the lower part. The structure of the upper band of the phonon density 
of states $\rho(\omega)$ is clearly influenced by the
contributions of the bulk-like vibrations of the inner iron layers,
while the wide background is added by the interface atomic layers.

\begin{figure}[t!] \centering 
\includegraphics[width=8cm]{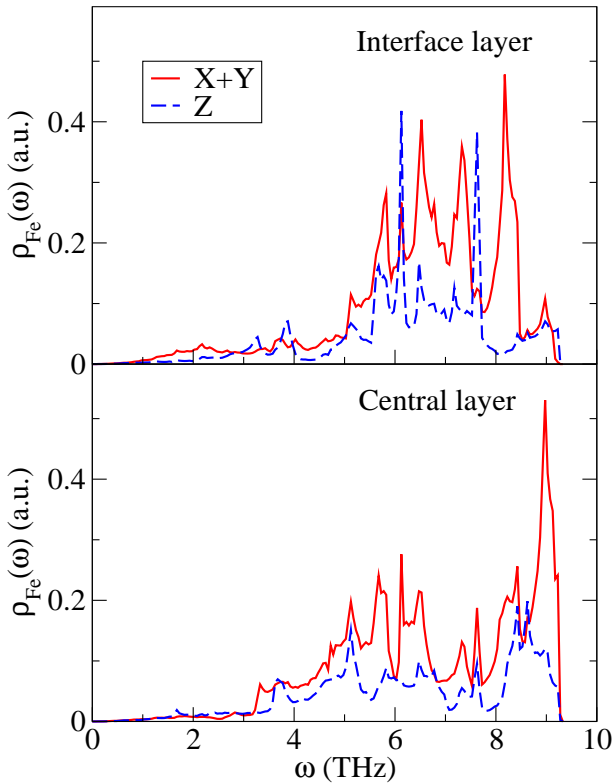}
\caption{
Partial phonon densities of states $\rho_{\rm Fe}(\omega)$ projected 
at Fe atoms of the 4/4 Fe/Pt multilayer structure in FM configuration 
for: interface (top), and central (bottom) Fe ML. In each panel the 
densities of states for $X$--$Y$ and $Z$ polarizations are 
plotted separately by solid (red) and dashed (blue) lines, respectively.}
\label{fig:phdoslayer}
\end{figure}

While an experimental verification of the theory concerning the phonon 
dispersions in the Fe/Pt multilayers (Fig.~\ref{fig:dispcurv}) is not 
possible at the moment, we suggest that the projected phonon densities of 
states at iron atoms $\rho_{\rm Fe}(\omega)$ could be measured by the 
nuclear inelastic scattering, which make use of the M\"ossbauer effect
\cite{Roh01}. Since the 
M\"ossbauer experimental techniques are sensitive only to iron atoms 
and depend on polarization for the considered type of layered structure, 
we extracted the projected phonon densities of states for two different 
iron MLs in the 4/4 structure, with $X-Y$ planar vibrations, and with 
perpendicular $Z$ vibrations, see Fig.~\ref{fig:phdoslayer}. 
As the projected phonon density of states for non-equivalent inner iron 
MLs of the 5/5 system are rather similar to each other, we analyse in 
detail $\rho_{\rm Fe}(\omega)$ obtained for the 4/4 system, as it shows 
already the general trends representative of the considered structures.

The differences between the $X-Y$ and $Z$ phonon polarizations in 
Fig.~\ref{fig:phdoslayer} are remarkable. While the spectral weight for the 
$Z$ polarization is more uniformly distributed, one finds characteristic 
maxima in the $X-Y$ density of states near $\omega\simeq 6.5$ THz and 
$\omega\simeq 8.8$ THz for the central layer (bottom of 
Fig.~\ref{fig:phdoslayer}). At the interface (top of 
Fig.~\ref{fig:phdoslayer}) an additional structure at $\omega\simeq 7.2$ 
THz was identified, while the maximum at high energy is moved down to 
$\omega\simeq 8.0$ THz, with low spectral weight above $8.3$ THz. 
Also for the $Z$ polarization the spectral intensity in the high 
energy part $\omega>8$ THz is reduced. 

Altogether, the phonon spectra show the quasi-two-dimensional character 
of the investigated multilayer structure, with large spectral weight in 
the range of $5.5<\omega<8.5$ THz for the interface, and a broader and 
extending up to $\sim 9.3$ THz projected phonon density of states 
$\rho_{\rm Fe}(\omega)$ for the inner Fe ML (Fig.~\ref{fig:phdoslayer}). 
This feature follows from the coupling to heavier Pt atoms at the 
interface, so the phonon frequencies are there reduced. The opposite 
situation was found for Fe/FeSi interface \cite{Joc06}, where the 
spectral weight was shifted to higher frequency range for the modes which 
involve light Si atoms at the interface. However, the projected phonon 
density of states found for the inner Fe ML in Fe/Pt system has several 
features (overall distribution of spectral weight, the positions of maxima) 
which make it similar to that of the inner Fe ML in Fe/FeSi multilayer. 
This confirms that the lattice vibrations depend predominantly on the local 
coordination of particular Fe atoms, which follows from the short-range 
character of interatomic forces.

\section{Summary and Conclusions}
\label{conclusions}

We have calculated the magnetic properties and lattice dynamics of 
the set of Fe/Pt multilayer systems of varying thickness. In order 
to demonstrate the universal stability of the FM configuration between
neighbouring Fe layers, we investigated spacer thickness dependence of 
the energy difference between the AF and FM structures. This energy 
difference was quantified using the bilinear ($J_1$) and biquadratic 
($J_2$) coupling coefficients. It is quite remarkable that the values of 
$J_1$ and $J_2$ saturate fast with increasing thickness of Pt layer, 
indicating that the FM order of consecutive Fe layers is hardly 
influenced by the Pt spacer. These data may be therefore considered 
to be representative of larger systems as well.

A systematic analysis of the magnetic structures shows that the 
magnetic properties of Fe/Pt multilayer are quite different from those 
of Fe/FeSi multilayers, where a crossover from the AF to FM interlayer 
coupling (between Fe layers) was found instead, with increasing thickness 
of the FeSi spacer layer \cite{Joc06}. In the case of Pt spacer 
the tendency towards FM order is well pronounced, and the magnetic 
moments of Fe atoms at the interface are enhanced. This agrees with 
earlier experimental findings which revealed rather robust tendency 
towards FM order in the Fe/Pt multilayers \cite{Sim96}.

The phonon study has demonstrated that the phonon spectra consist of 
several modes, reflecting the geometry of a considered system with 
large unit cell. Platinum and iron vibrations are rather well 
separated from each other, and the vibrations of iron atoms contribute 
predominantly at higher energies $\omega>3.5$ THz ($\omega>5$ THz in 
case of the interface layer), as shown by the respective partial phonon 
densities of states. The phonon spectra projected on iron atoms depend 
rather strongly on the multilayer geometry, and differ distinctly from 
those of the bulk Fe \cite{Laz06}, both for the interface and for the
inner Fe MLs, at least when the iron layer consists of not more than 
five atomic MLs, as considered in the present study. We believe that the 
maxima in the partial phonon density of states projected on iron atoms 
shown in Fig.~\ref{fig:phdoslayer} should be fairly easy to observe and 
verify experimentally. Therefore, we hope that the present theoretical 
predictions concerning lattice dynamics will help to understand the 
intricate dependence of the phonon spectra on the multilayer structure. 

Summarizing, the Fe/Pt multilayers exhibit not only interesting 
magnetic properties, but have also non-trivial and rich lattice 
dynamics, characterized by rather strong dependence of the partial 
phonon intensities on the position of an Fe monolayer in the structure. 
It is believed that the presented results are representative for the 
Fe/Pt systems and will stimulate future experiments.

\section{acknowledgments}

This work was partially supported by the European Community under 
FP6 contract No. NMP4-CT-2003-001516 (DYNASYNC). A.M. Ole\'s
acknowledges support by the Foundation for Polish Science (FNP).

\end{document}